\documentclass[aps,pre,twocolumn,nofootinbib]{revtex4-1}
\usepackage{amsmath,amssymb,amstext}
\begin{document}
\title{Relativistic corrections to the central force problem in a generalized potential approach}
\author{Ashmeet Singh}
\email{ashmtuph@iitr.ac.in}
\affiliation{Department of Physics, Indian Institute of Technology Roorkee, India - 247667}
\author{Binoy Krishna Patra}
\email{binoyfph@iitr.ac.in}
\affiliation{Department of Physics, Indian Institute of Technology Roorkee, India - 247667}
\newcommand{\lag}{\mathcal{L}}
\newcommand{\ham}{\mathcal{H}}
\begin{abstract}
We present a novel technique to obtain relativistic corrections
to the central force problem in the Lagrangian formulation, using a 
generalized potential energy function. We derive a general expression for a generalized potential energy function for all powers of the velocity, which when made a part of the regular classical Lagrangian can reproduce the correct (relativistic) force equation. We then go on to derive the
Hamiltonian and estimate the corrections to the 
total energy of the system up to the  fourth power in $|\vec{v}|/c$. We found that our work is able to provide a more comprehensive understanding of relativistic corrections to the central force results and provides corrections to both the kinetic and potential energy of the system.
 We employ our methodology to calculate relativistic corrections to the circular orbit under the
gravitational force and also the first-order 
corrections to the ground state energy of the hydrogen atom 
using a semi-classical approach. Our predictions in both problems give
reasonable agreement with the known results.  Thus we feel that 
this work has pedagogical value and can be 
used by undergraduate students to better understand the central 
force and the relativistic corrections to it.
\end{abstract}
\maketitle
\section{Introduction} 
The central force problem, in which the ordinary potential function 
depends only on the magnitude of the relative separation between the 
particles $(r)$ is a very well studied problem in classical and quantum 
mechanics. The problem is particularly useful since central 
potentials are common in physics, e.g. the electrostatic Coulomb 
and the Newtonian gravitational potential. The two-body problem under the central potential 
is solved in any standard textbook of classical mechanics (eg. \cite{ranaJoagCentral}) 
to obtain the equation of orbit; the hydrogen atom\cite{griffithsHydrogen} problem in quantum mechanics, to yield expectation values 
of physical observables, etc. In particular, the total angular 
momentum, the total mechanical energy of the 
system are constants of motion.\footnote{The Runge-Lenz tensor is also a constant of motion.} The standard approach in 
non-relativistic classical mechanics is to convert the two-body 
central force equation into an equivalent one-body problem.\cite{tukwalePuranikOneBody}
The classical Lagrangian for this holonomic system  and the  corresponding Jacobi Integral of this reduced system 
can be easily written, followed by solving the equation of orbit. 
The Lagrangian is identified as $\lag = T - V$, and 
the Hamiltonian or Jacobi Integral as $\ham = T + V$, with $T$ 
being the kinetic energy and $V$ the ordinary potential energy functions.

The Lagrangian and Hamiltonian formulations can be made 
compliant with a relativistic framework, which gives rise
 to the correct spatial equations of motion and the total 
energy as observed from a given inertial reference frame.
 Covariant Lagrangian and Hamiltonian formulations\cite{covariant} 
exist  which treat space and time on a common footing as the 
generalized coordinates in a four-dimensional configuration 
space. This is indeed the correct relativistic formulation 
which is Lorentz invariant, but it is seen that the 
complexity of the equations is an issue, even for the 
simplest possible cases like that of a single particle. 
The true relativistic equation for the hydrogen atom is 
governed by the famous Dirac equation\cite{diracEq} in relativistic
quantum mechanics which offers excellent insight into 
the problem.
In this paper, we have incorporated relativistic corrections 
to the central force problem by using a generalized velocity 
dependent potential energy setup. Our Lagrangian is of the form, 
$\lag = T - U$ with $T$ being the non-relativistic kinetic energy and
$U$ being the generalized potential energy function.
The new Lagrangian constructed from the generalized potential energy function produces the correct relativistic equations of motion 
under the influence of central force. The Jacobi Integral or the Hamiltonian 
is then calculated from this modified Lagrangian using the 
standard prescription. While understanding the physics of 
these relativistic corrections, we see that our methodology provides a 
unified approach of the relativistic corrections to the 
(non-relativistic) central force. We employ our methodology 
to some physical problems, such as 
the ground state of the hydrogen atom\cite{resnickEisbergSpeed} 
semi-classically and orbits of celestial bodies. In these cases, particle speeds are semi-relativistic 
such that $| \vec{v} | <  0.1 c$ and this permits us to retain maximally 
upto quartic power of $|\vec{v}|/c$, since higher powers have negligibly 
small contributions. This approach is interesting since 
we are able to avoid the non-trivial complexities in the 
covariant formulation and are able to successfully add
relativistic corrections to the problem using the standard textbook 
approach. \\

A brief outline of the organization of the paper is as 
follows. In section 2, we add relativistic corrections 
to the attractive, inverse square central force to define a generalized force, which we use in 
section 3 to derive the generalized velocity dependent 
potential energy function. Section 4 develops the Lagrangian 
and Hamiltonian of the system, using the generalized potential energy function 
of Section 3, and calculates the total energy of the system. 
In section 5, we show two model applications 
of our methodology, to the classical gravitationally bound, circular 
orbits and the other to find the first-order correction to the 
ground state energy of the hydrogen atom. We conclude in section 6 by summarizing our work. 
\section{Relativistically Generalized Central Force} 
A central force acts along the position vector of a particle drawn
 from the center of force and depends only on the scalar 
distance $r = |\vec{r}|$ from the fixed center of force. The convenient
 choice of this fixed point to be the origin of the coordinate
 system makes the central force look like,
 \begin{equation}
 \label{centralForce}
\vec{F}(\vec{r}) = F(r) \hat{r} \: .
 \end{equation}
In case of the non-relativistic central force, the force is derivable
 from an ordinary potential energy function $V(r)$ such that,
\begin{equation}
\label{gradV}
\vec{F}(\vec{r}) \: = F(r) \hat{r} \:  = \: - \vec{\nabla} {V(r)} \: .
\end{equation}
Consider two point particles of rest masses $m_{1}$ and $m_{2}$ having
 a position vector $\vec{r}_{1}$ and $\vec{r}_{2}$, respectively, moving
 under the influence of central force field, where the equation of motions 
can be decomposed into the relative $\vec{r} (= \vec{r}_{1} - \vec{r}_{2}$) and
center-of-mass ($\vec{R}$) coordinates, respectively. In the absence of
any external force on this system, the center-of-mass ($\vec{R}$) of the system moves as a free
 particle. Thus in classical, non-relativistic mechanics, the two-body problem can be reduced to the dynamics of an
 equivalent one-body of a hypothetical mass point of reduced 
mass $\mu = ({m_{1} m_{2}})/({m_{1} + m_{2}})$ with the relative position 
vector $\vec{r}$ from the origin, moving under the action of the internal interaction force\footnote{Reduced mass does not have a simple definition in relativistic theory. However, we could treat the heavier mass as a source of static field and also the motion of the heavier mass can be taken as non-relativistic to a very good approximation\cite{reducedMass}. This justifies the use of an equivalent one-body having a reduced mass $\mu$.}.
In general, the relativistic force equation for a particle of rest 
mass $m_{0}$ in classical mechanics can be written as (Einstein summation implied),
\begin{equation}
\label{ForceEq}
F_{i} = m_{ij} \frac{d^{2} x_{j}}{dt^{2}} \: ,
\end{equation}
with $m_{ij}$ being the relativistic mass tensor,\cite{ranaJoagMassTensor}
\begin{equation}
\label{massTensor}
m_{ij}  = m_{0} \gamma ^{3} \left[ \frac{\delta _{ij}}{\gamma ^{2}}  + \frac{v_{i} v_{j}}{c^{2}} \right] \: .
\end{equation}
In the above equation, symbols have their usual meanings 
with $v_{i}$ being the $i^{\rm{th}}$ component of the spatial velocity 
and $\gamma$ being the 
Lorentz Factor $\gamma = \left(1 - \frac{|\vec{v}|^{2}}{c^{2}} \right) ^{-1/2}$ and $\delta _{ij}$ the Kronecker's delta. To understand the relative importance of the two terms in the mass tensor (Eq. \ref{massTensor}) to the total force, we rewrite the force equation of Eq. (\ref{ForceEq}) in its vector notation,
\begin{equation}
\label{ForceVector}
\vec{F} = \frac{d}{dt}(\gamma m_{0} \vec{v}) = m_{0} \left( \gamma \frac{d^{2} \vec{r}}{dt^{2}} + \frac{\gamma^{3}}{c^{2}} (\vec{v} \cdot \vec{a}) \vec{v}
\right)		\:,
\end{equation}
where the first term in the acceleration $\left(\gamma 
{d^{2}\vec{r}}/{dt^{2}} \right)$ 
is the contribution of the diagonal term of the mass tensor 
and the second term $\left(\gamma^{3} (\vec{v} \cdot \vec{a}) 
\vec{v}/ {c^{2}} \right)$ represents the off-diagonal contribution
which is a product of two terms:  $\left( \gamma^{3} \vec{v}/c^{2} 
\right)$ and $\left( \vec{v} \cdot \vec{a} \right)$.
The term $\left( \gamma^3~\vec{v}/{c^2} \right)$ is much smaller than unity since 
$v/c < 1$. The term ($\vec{v} \cdot \vec{a}$) is also very small - for elliptic orbits with small eccentricities, the velocity is nearly perpendicular to the acceleration and hence $\vec{v} \cdot \vec{a}$ is close to zero. It becomes
identically zero for the circular orbits, where the velocity is perpendicular to the acceleration.
We can thus safely argue that the off-diagonal term in the 
acceleration is negligibly small compared to the diagonal term. 
Objects in our Solar System orbiting the Sun have very 
low eccentricities. For example, the Earth's orbit is nearly circular with
a small eccentricity of 0.017\cite{earthE} and Pluto is the body with the 
largest mean eccentricity of 0.244\cite{earthE}. A simple calculation 
for the orbit of Pluto yields that the off-diagonal contribution of
the force is $\sim \mathcal{O}{\left(10^{-17}\right)}$ smaller than 
the diagonal term. Also, on the other extreme (in the microscopic 
length scale), the 
relativistic Sommerfeld analysis\cite{sommerfeld} predicts the low eccentricity
for the elliptical orbits of the electron in the hydrogen atom. 
Thus, for an attractive central force,
\begin{equation}
\vec{F}(r) = {-k}/{r^{2}}~\hat{r} ~~~~~~~~~~k>0~~,
\end{equation}
the force equation (5) for the reduced one-body problem, after neglecting 
the non-diagonal term, can therefore be written  as
\begin{equation}
\label{ForceEq4}
\mu \ddot{\vec{r}} = {{\cal{Q}}} ~\hat{r}\: ,
\end{equation}
where ${\cal{Q}}$ is the generalized force, and its magnitude is given by
\begin{equation}
{\cal{Q}} = -\frac{k}{r^{2}} \left( 1 - \frac{|\dot{\vec{r}}|^{2}}{c^{2}} 
\right)^{1/2}~,
\end{equation}
which depends on both the relative separation ($r$) and the 
velocity ($\dot{\vec{r}}$) of the particle. This re-arrangement allows us 
to write the force equation as typical 
Newtonian equation of motion and use the familiar non-relativistic 
Lagrangian Formulation with $\lag = T - U$, ($U$ being 
the generalized velocity dependent potential energy function) such
that the Euler-Lagrange differential equation of motion\cite{ranaJoagELDE}
can reproduce the correct force equation.

Under the central force, the motion is planar and we can
use two-dimensional polar coordinates $(r, \theta)$ to describe
the motion of the reduced particle. Thus, the velocity of the reduced
particle can be resolved into its radial and tangential components 
as,
\begin{equation}
\label{velBreak}
 |\dot{\vec{r}}|^{2} \: = \: {\dot{r}}^{2} + r^{2} {\dot{\theta}}^{2} \: .
\end{equation} 
Therefore the force equation (\ref{ForceEq4}) can be be decomposed into 
radial and tangential components:
\begin{equation}
\label{ForceR}
\mu \ddot{r} - \mu r {\dot{\theta}}^{2} = {\cal{Q}}
\end{equation}
\begin{equation}
	\label{Lconstant}
\mu r \ddot{\theta} + 2 \mu \dot{r} \dot{\theta} = 0 \: . 
\end{equation}
We can clearly see from Eq. (\ref{Lconstant}) that the total angular momentum
 $L$ is a constant of motion such that,
\begin{equation}
\label{angMom}
\frac{dL}{dt} = \frac{d}{dt} (\mu r^{2} \dot{\theta}) = 0 \: .
\end{equation}
Hence, we can eliminate $\dot{\theta}$ in terms of $L$ and $r$, so that the planar problem
again be reduced to the radial ($r$) problem only. 

For $ |{\vec{v}}| / c <1$, ($\vec{v}=\dot{\vec{r}}$), 
we can write the generalized force by expanding the 
Lorentz factor in powers of $|{\vec{v}}|/c$ using a binomial expansion,
\begin{equation}
\label{expForce_2}
{\cal{Q}} \: =  -\frac{k}{r^2} \left[ \sum_{p = 0}^{\infty} {\frac{1}{2}\choose{p}} \left(\frac{|\vec{v}|}{c}\right)^{2p} \right] 	\:	,
\end{equation}
where ${\frac{1}{2}\choose{p}}$ is the binomial coefficient of 1/2 and $p$. 
Due to the conservation of angular momentum, the velocity $|\vec{v}|$
becomes
\begin{equation}
\label{vel2}
|\vec{v}| = \frac{L^2}{\mu^2r^2} \left( 1+ \frac{\mu^2 r^2}{L^2} {\dot{r}}^2
\right)~,
\end{equation} 
so we further expand $\left({|\vec{v}|}/{c}\right)^{2p}$ using Eq. (\ref{vel2}) in Eq. (\ref{expForce_2})
in terms of a power series in the radial speed $\dot{r}$ only. Therefore
the generalized force becomes
\begin{equation}
\label{finalFcorrSeries}
{\cal{Q}} \: =  k \left[ \sum_{p = 0}^{\infty} \sum_{q = 0}^{p} B_{p,q}(L,\mu)  r ^{2(q-p-i)} (\dot{r}^{2q}) \right] 	\:	,
\end{equation}
where we have defined $B_{p,q} (L, \mu)$ as the expansion coefficient depending on $L$ and $\mu$  given by,
\begin{equation}
\label{Ycoeff}
B_{p,q} (L,\mu) \: = \frac{(-1)^{p+1}}{c^{2p}} {\frac{1}{2}\choose{p}} {p\choose{q}}  \left(\frac{L}{\mu} \right)^{2(p-q)} \: 	.
\end{equation}
It is proper to mention here that the generalized force for a relativistic 
particle of rest mass $\mu$ contains all powers in $|\vec{v}|/c$. 
In this paper, we have limited ourselves upto the quartic power in 
$|\vec{v}|/c$ since higher order terms will have negligibly small 
contribution in the semi-relativistic regime. 
Interested readers are encouraged to explore the contribution of higher 
powers to the central force.  Thus, up to fourth power in $|\dot{\vec{r}}|/c$,
the generalized force becomes
\begin{equation}
\label{expForceFinal_pre}
{\cal{Q}} = -\frac{k}{r^{2}} + \frac{k L^{2}}{2 \mu^{2} c^{2} r^{4}} 
+ \frac{k L^{4}}{8 \mu^{4} c^{4} r^{6}} + 
\frac{k {\dot{r}}^{2}}{2 c^{2} r^{2}} +  
\frac{k L^{2} {\dot{r}}^{2}}{4 \mu^{2} c^{4} r^{4}} 
+\frac{1}{8}\frac{k}{c^4r^2} {\dot{r}}^{4} \: ,
\end{equation}
where the first term is the usual inverse square attractive force. 
Both the second and third terms represent the relativistic corrections to 
the (non-relativistic) central force and allow for the coupling of the 
angular momentum component with the central force component. However, in 
the third term higher powers of angular momentum couples with the central force.
Later we will see that these two terms in the generalized force will 
contribute to the (relativistic) corrections of the potential energy of 
the system. The fourth and fifth terms which depend on the square of the radial 
speed and also couple the angular momentum of the system with the 
central force, will represent quadratic and quartic power corrections 
to the kinetic energy of the system respectively. The sixth term is a higher order kinetic correction to the central force of the system depending on the fourth power of the radial speed. \\ \\
We add a further simplification here - building on the fact that $(\vec{v} 
\cdot \vec{a})$ (in Eq. \ref{ForceVector}) is negligibly small for closed 
orbits of smaller eccentricity. It can be shown that 
the term $(\vec{v} \cdot \vec{a})$, which is a measure of the eccentricity, 
is proportional to the radial velocity, $\dot{r}$, 
so for a circular orbit, $\dot{r} = 0$ identically, and for nearly circular orbits, $\dot{r}$ is small. Out of the three terms in the expansion having 
fourth power in $|\dot{\vec{r}}|/c$ in the generalized force, we neglect only 
the term having the fourth power of the radial speed $\sim \dot{r}^4/c^4$ 
but retain the contribution of the fourth power 
of tangential speed and a mixed biquadratic term in radial and
tangential velocities. Thus the generalized force of Eq. (\ref{expForceFinal_pre}) is simplified into 
\begin{equation}
\label{expForceFinal}
{\cal{Q}} = -\frac{k}{r^{2}} + \frac{k L^{2}}{2 \mu^{2} c^{2} r^{4}} 
+ \frac{k L^{4}}{8 \mu^{4} c^{4} r^{6}} + 
\frac{k {\dot{r}}^{2}}{2 c^{2} r^{2}} +  
\frac{k L^{2} {\dot{r}}^{2}}{4 \mu^{2} c^{4} r^{4}} \: ,
\end{equation}
which will make the radial momentum (Eq. \ref{pr}) linear in radial 
velocity, $\dot{r}$ and presents a welcome simplification 
to construct the Hamiltonian from the Lagrangian without 
loosing much on the physics of the problem.

\section{The Relativistic Generalized Potential} \label{Ucorrect}
We now construct a Lagrangian function $\lag (r , \dot{\vec{r}} ,t)$ of
 the non-relativistic, standard form of $\lag = T - U$, 
such that Eq. (\ref{ForceR}) can be reproduced through the Euler-Lagrange equation of motion satisfying,
\begin{equation}
\label{elde}
\frac{d}{dt} \left(\frac{\partial \lag}{\partial \dot{r}} \right) - \frac{\partial \lag}{\partial r} = 0 \: ,
\end{equation}
and the right hand side of the above equation is zero since there are no non-potential forces acting on the system. 
The kinetic energy function $T$ is chosen to have the Newtonian, non-relativistic form,
\begin{equation}
\label{Tfunc}
 T(r, \dot{r}, t) = \frac{1}{2} \mu |\dot{\vec{r}}|^{2} \: ,
\end{equation}
and $U \equiv U(\vec{r}, \dot{\vec{r}},t)$ is
 the generalized potential energy function. The potential energy function $U$ will be
 chosen such that the generalized force $\cal{Q}$ should be
 derivable from $U$ as,
\begin{equation}
\label{ForcePotential}
{\cal{Q}} = \frac{d}{dt} \left(\frac{\partial U}{\partial \dot{r}} \right) - \frac{\partial U}{\partial r} \: .
\end{equation}
Since the corrected force also depends on the radial velocity $\dot{r}$ in
 addition to $r$, hence an ordinary potential $V(r) = -k/r$ will not be sufficient
 to describe the generalized force. At this stage, we can expect that the
 generalized potential $U$ to have a dominant contribution from the
 ordinary potential term $V$ and other additive, corrective terms which
 depend on $r$ and $\dot{\vec{r}}$. Since the problem
 is entirely in one-dimension in the $r$ coordinate, we label without 
confusion $\dot{r} \equiv v$.
Since the relativistically generalized force expansion of Eq. (\ref{finalFcorrSeries}) contains only even powers of $\dot{r}$, we choose a power series  for the generalized potential $U(r,v,t)$ in powers of $v^2$, with the expansion coefficients
 depending on $r$ and $t$,
\begin{equation}
\label{Useries}
U(r,v,t) = \sum_{n = 0}^{\infty} h_{2n} (r,t) \: v^{2n} 	\:	,
\end{equation}
where in the above equation, $h_{2n} (r,t)$ is the 2n-th order expansion coefficient. This power series for $U$ is used to find an expression for the generalized force,
\begin{equation}
\label{ForceSeries}
{\cal{Q}} = \sum_{n=0}^{\infty} {(2n-1) \frac{\partial h_{2n}}{\partial r} \: v^{2n} } +  2n(2n-1) h_{2n} \: v^{2n-2} \dot{v} \: ,
\end{equation}
where we have assumed that the expansion coefficients are independent of time explicitly. This allows welcome simplification and is justified since there is no
 explicit time dependence in the generalized force equation as well. 
The generalized force $\cal{Q}$ of Eq. (\ref{finalFcorrSeries}) 
is independent of the particle's radial acceleration $\dot{v}$, as is the case with most naturally occurring forces in nature
 and hence the sum of the second term in Eq. (\ref{ForceSeries}) must result up to zero, we get,
\begin{equation}
\label{noAcc}
\sum_{n=0}^{\infty} (2n)(2n-1) h_{2n}(r) \: v^{2n-2} = 0 \: ,
\end{equation}
and thus, the generalized force can be expressed in a power series form as follows,
\begin{equation}
\label{forceSeriesFinal}
{\cal{Q}} = \sum_{n=0}^{\infty} (2n-1) \frac{\partial h_{2n}(r)}{\partial r} \: v^{2n}  \: .
\end{equation}
We now compare the coefficients in Eq. (\ref{forceSeriesFinal}) 
with the terms of the force of Eq. (\ref{finalFcorrSeries}) in 
equal powers of $v$, which gives us a general expression for $h_{2n} (r)$,
\begin{equation}
\label{h2nCoeff}
h_{2n} (r) \: = \frac{k}{2n-1} \sum_{p=n}^{\infty} B_{p,n}(L,\mu) \frac{r^{2(n-p) - 1}}{2(n-p) - 1} + A_{2n}	\:	,
\end{equation}
with $A_{2n}$ being the constants of integration, to be determined by boundary conditions. Thus, we have constructed a general expression for the generalized potential $U(r,v)$ which when plugged in Eq. (\ref{ForcePotential}) will give the exact expansion for the corrected force. To determine the integration constants $A_{2n}$, we use the usual
 boundary condition as employed in standard analysis of the central force.
 As $r \rightarrow \infty$, the force field dies down to zero. In this limit, the potential goes to zero and the first term ($r$-dependent) in the expansion coefficients $h_{2n}(r) = 0$ vanishes, we get,
\begin{equation}
\label{Ubc}
\lim_{r \to \infty} U(r,v) = 0 = \sum_{n=0}^{\infty} A_{2n} v^{2n} \: .
\end{equation}
Since this holds for any arbitrary radial velocity
 $v$ of the particle, we find all $A_{2n}$ to be zero identically, for all $n$,
\begin{equation}
\label{constants}
A_{2n} =  0 \: .
\end{equation}
By comparing with the relativistic force equation Eq. (\ref{expForceFinal}), we obtain 
the expansion coefficients as
\begin{equation}
\label{h0}
h_{0} (r) = -\frac{k}{r} + \frac{k L^{2}}{6 \mu ^{2} c^{2} r^{3}} + \frac{k L^{4}}{40 \mu ^{4} c^{4} r^{5}} + A_{0} \: ,
\end{equation}
\begin{equation}
\label{n1map}
h_{2} (r) = -\frac{k}{2 c^{2} r} - \frac{kL^2}{12 \mu^{2} c^{4} r^{3}} + A_{2} \:  .
\end{equation}
Thus, upto fourth power in $|\vec{v}|/c$, the generalized potential energy can be written as,
\begin{equation}
\label{U4th}
U(r,v) = -\frac{k}{r} + \frac{k L^{2}}{6 \mu ^{2} c^{2} r^{3}} + \frac{k L^{4}}{40 \mu ^{4} c^{4} r^{5}} - \frac{k v^2}{2 c^{2} r} - \frac{kL^2 v^2}{12 \mu^{2} c^{4} r^{3}} \: .
\end{equation}
We see an ordinary potential correction coming from $h_{0}(r)$ which tends to couple the angular momentum with the central force constant $k$. There are additional ``kinetic" corrections proportional to the square of the radial velocity arising from $h_{2}(r)$ and these will modify the kinetic energy of the system, as we shall see later. The contribution of these terms will be discussed in detail in the next section.
\section{Lagrangian and Hamiltonian Formulations}
\label{lagHam}
We now proceed to build a Lagrangian and following which, 
a Hamiltonian of the relativistic system under central force.
 As suggested earlier, we will use a Lagrangian function $\lag (\vec{r} , \dot{\vec{r}} , t)$ ( $  = T - U$), with the relativistically 
corrected generalized potential $U$ found in the previous section. 
With such a form, we see that the Euler-Lagrange equation will reproduce the 
relativistic force equation. 
Thus, we now construct the Lagrangian as,
\begin{equation}
\label{lag1}
\lag(\vec{r}, \dot{\vec{r}}) = \: \frac{1}{2} \mu v^{2} + 
\frac{L^{2}}{2 \mu r^{2}} - U(r,v) \:
\end{equation}
where $U(r,v)$ can be substituted from Eq. (\ref{Useries}) and 
the coefficients from Eq. (\ref{h2nCoeff}).
Working only till fourth power in $|\vec{v}|/c$, we can utilize the potential from Eq. (\ref{U4th}). Once the Lagrangian
 has been constructed, the canonical momentum conjugate to the radial
 coordinate $r$ can be calculated as follows,
\begin{equation}
\label{pr}
p_{r} = \frac{\partial \lag}{\partial \dot{r}} = \frac{\partial \lag}{\partial v} = \left( \mu + \frac{k}{c^{2} r} + \frac{kL^2}{6 \mu^2 c^4 r^3} \right)v \: .
\end{equation}
It is here we see the usefulness of neglecting the term proportional to fourth power in radial speed in Eq. (\ref{expForceFinal}). It allows us to keep the radial momentum linear in the radial speed. Relativistic corrections can be clubbed into a single pre-factor which corrects the non-relativistic momentum $\mu v$. This is in line with our motivation to use a Newtonian framework for formulating relativistic corrections to the central force.
Thus the canonical momentum can be expressed in a succinct form 
by allowing comparison with the classical, non-relativistic momentum as,
\begin{equation}
\label{pr_compare}
p_{r}  = \left[1 + \frac{\Gamma}{r} \left(1 + \frac{L^2}{6 \mu^2 c^2 r^2} \right)  \right] \mu v	  = \chi (r) \mu v \: .
\end{equation}
In the above equation, we have substituted $\Gamma$ for ${k}/{(\mu c^{2})}$, 
which is an interaction parameter, useful in better understanding the physics 
of the problem. 
The parameter $\Gamma$ is an indicative of the relative strength of the
 central force coupling to the rest mass energy $\mu c^{2}$ of the system,
 and has dimensions of length. It allows us to compare the rest mass energy with the binding energy of the system. In the limit $\Gamma << 1$,
 the problem can be treated non-relativistically since the rest mass energy 
is much larger the binding energy of the system, while in the opposite limit, 
the treatment is required to be relativistic, since the binding energy is 
comparable to the rest mass energy of the system. 
In many physical situations, $\mu c^2$ is very large compared to the binding energy of 
the system, for example, an electron in the hydrogen atom, the rest mass energy 
($\sim 511 \rm {keV}$) is much larger than its binding energy ($\sim 10 
\rm {eV}$) so the ground state properties can be understood
through the non-relativistic Schr\"{o}dinger wave equation. The entire 
pre-factor has been labeled as $\chi (r)$ to allow expressions to 
have a condensed form.
The angular momentum $p_{\theta}$, canonically conjugate
 to the $\theta$ coordinate is a constant of motion as we had seen
 earlier and labeled it as $L$. The Hamiltonian $\ham$ of the
 system can now be calculated as a Legendre dual transform
 of the Lagrangian,\cite{ranaJoagHamiltonian}
\begin{equation}
\label{ham2}
\ham (\vec{r},\vec{p},t) = p_{r} \dot{r} + p_{\theta} \dot{\theta} - \lag \: .
\end{equation}
Again eliminating $\dot{\theta}$ using the fact that the
 angular momentum $p_{\theta} \equiv L$ is a constant of motion,
 we can express the Hamiltonian in terms of the radial coordinate
 and its conjugate momentum exclusively,
\begin{equation}
\label{energy}
\ham = \frac{p^{2}_{r}}{2 \mu \chi(r)} + \frac{L^{2}}{2 \mu r^{2}} - \frac{k}{r} +\frac{\Gamma L^2}{6 \mu r^3} + \frac{\Gamma L^4}{40 \mu^3 c^2 r^5}  	\:	.
\end{equation}
Since the Hamiltonian does not depend on time explicitly, so it can be identified as the total energy $E$ of the system, which is a constant of motion.
Thus, the total mechanical energy
 of a system is seen to have relativistic correction to both the kinetic and 
potential energy terms. It can readily be verified that in the 
$\Gamma \rightarrow 0$ limit, we recover the 
non-relativistic results. 
In the next section, we present two model applications of 
this relativistically corrected central force formulation to physical systems under the influence of the central force.
\section{Model Applications} \label{application}
We now present model applications of the relativistic central force
 formulation using a generalized potential to the classical circular orbit in celestial mechanics under the gravitational force and the 
first order energy correction to the ground state of the hydrogen atom in quantum mechanics. 
These applications can be used as a pedagogical tool in the 
undergraduate classroom to understand the implications of 
relativistic corrections to the central force.
These can be seen as approximations that bring out the 
physics of the problem without having to indulge non-trivial mathematics.
\subsection{Relativistic Correction to the Classical Circular Orbit}
\label{orbit}
We shall now be considering a gravitationally bound circular orbit
 and calculate a relativistic correction to it. Before 
we calculate the correction, we assert that in case of planetary orbits in 
the solar system, $|\dot{\vec{r}}|/c < 10^{-4}$, so we retain only in quadratic 
powers in $|\dot{\vec{r}}|/c$. This would also allow us to give a closed 
form to the relativistic correction to the radius of the circular orbit. 
In this case,
 the central force constant $k = G m_{1} m_{2}$, where $G$ is
 the universal gravitational constant. As is customary
 in non-relativistic central force problems, we define an
 effective potential $V_{\rm {{\rm{eff}}, {\rm{NR}}}}$  (${\rm{NR}}$ 
stands for non-relativistic), which compares the
 effect of the (attractive) central potential and the (repulsive) centrifugal 
terms\cite{Veff_TP}, as 
\begin{equation}
\label{veffNR}
V_{\rm {eff, NR}} (r) = \frac{-G m_{1} m_{2}}{r} + \frac{L^{2}}{2 \mu r^{2}} 	\: 	.
\end{equation}
On similar lines, we can define a relativistic effective potential
 for the central force which has an additional corrective term,
\begin{equation}
\label{veffR}
V_{\rm {eff, rel}} (r) = \frac{-G m_{1} m_{2}}{r} + \frac{L^{2}}{2 \mu r^{2}} 	+ \frac{\Gamma L^{2}}{6 \mu r^{3}} \: 	,
\end{equation}
\begin{equation}
\label{veffR_2}
V_{\rm {eff, rel}} (r) =   V_{\rm {eff, NR}} (r) + \frac{\Gamma L^{2}}{6 \mu r^{3}} \: 	,
\end{equation}
such that the energy of the system can now be compactly written as (where `rel' stands for relativistic), in terms of effective
potential
\begin{equation}
\label{EinVeff}
E = \frac{1}{2}  \mu \left(1 + \frac{\Gamma}{r} \right) v^{2} + V_{\rm {eff, rel}}	\: .
\end{equation}
We can now solve for the radial speed $\dot{r}$ by re-arranging 
terms in Eq. (\ref{EinVeff}),
\begin{equation}
\label{rdot}
{\dot{r}}^{2} = \frac{2}{\mu} \left(1 + \frac{\Gamma}{r} \right) ^{-1} \left(E - V_{\rm {eff, rel}} (r) \right)	\:	.
\end{equation}
It is known that that bound orbits, under the classical central force 
problem of an inverse square attractive potential, occur when
 $\rm{min(V_{eff})} \leq E < 0$.\cite{Veff_TP} A special case occurs when
 $E = \rm {min(V_{ eff})}$ such that $\dot{r} = 0$ and the resultant orbit
 is circular. Non-relativistically, a circular orbit occurs 
when $E = \rm {min(V_{eff,NR})}$ at a radius $r_{0}$,
\begin{equation}
\label{r0}
r_{0} = \frac{L^{2}}{\mu k} \equiv 2 \eta	\:	,
\end{equation}
where, we have defined $\eta$ for mathematical convenience. 
Relativistically, we will now give an estimate of the radius of the
 circular orbit for a given value of $L$. We observe that \\
 $\left(1 + \frac{\Gamma}{r} \right) ^{-1} \neq 0$, hence, $\dot{r} = 0$ for a circular orbit occurs only when $E = \rm{min(V_{eff, rel})}$. 
Let the minimum of $\rm {V_{eff, rel}}$ occur at $r = r_{*}$, such that,
\begin{equation}
\label{min1}
\left( \frac{d  V_{\rm {eff, rel}}}{dr} \right) _{r_{*}} = 0 		\:  \rm{and} \: 
\left( \frac{d^{2}   V_{\rm {eff, rel}}}{d r^{2}} \right) _{r_{*}} > 0 \: . 	
\end{equation}
The stationarity condition of Eq.(\ref{min1}) can be
 solved to yield the following quadratic equation in $r$,
\begin{equation}
\label{rQuadratic}
r^{2} - 2 \eta r - \Gamma \eta = 0 	\:	,
\end{equation}
The radius of the relativistic circular orbit is found to be,
\begin{equation}
\label{rstar1}
r_{*} = \frac{r_{0}}{2} \left[ 1 + \left( 1 + \frac{\Gamma}{\eta} \right)^{1/2}	 \right] \equiv  r_{0} \delta 	\:	,
\end{equation} 
where we have defined $\delta$ as a parameter to
 compare the circular radii for the non-relativistic and
 relativistic cases, for a given $L$ and coupling $k$,
\begin{equation}
\delta = \frac{1}{2}  \left[ 1 + \left( 1 + \frac{2 k^{2}}{L^{2} c^{2}} \right)^{1/2}	 \right]		\:	.
\end{equation}
It is readily seen that $\delta > 1$, thus, the radius of
 the circular orbit with the relativistic correction is more
 than without it, $r_{*} > r_{0}$. The presence of the additional 
positive term in the relativistic effective potential, $\left(\Gamma L^2/6\mu
r^3 \right)$ tends to flatten the potential curve at larger radii and can be understood as the reason for the swelling of the circular orbit.
It can easily be calculated that the energy of the circular
 orbit with the relativistic correction $E_{*}$ is,
\begin{equation}
\label{Estar}
E_{*} =  \frac{E_{0}}{\delta} + \frac{\mu k^{2}}{2 L^{2}\delta} \left( \frac{1}{\delta} + \frac{k^{2}}{3 \delta^{2} L^{2} c^{2}} - 1 \right) \: 		,
\end{equation}
where $E_{0}$ is the energy of the non-relativistic circular
 orbit with the same angular momentum,
\begin{equation}
\label{E0}
E_{0} = V_{\rm {eff, NR}} (r_{0}) = \frac{-\mu k^{2}}{2 L^{2}}	\:	.
\end{equation}
Thus, we have calculated the radius and energy of a circular
 orbit under the given central force with the relativistic correction
 incorporated and have presented our results in a comparative way 
with the non-relativistic ones for a given value of $L$.
\subsection{Relativistic Correction to the Ground State Energy of the Hydrogen Atom in Vacuum}
In this section, we present a first order correction to the energy
 of the ground state of the hydrogen atom in vacuum due to
 relativistic corrections in the Hamiltonian. We assume that the
 proton is infinitely heavy as compared to the electron and can
 therefore be assumed to be stationary. Thus, the electron feels an electrostatic Coulomb force of attraction and in vacuum,
 $k = e^{2} / (4 \pi \epsilon _{0})$ with $e$ being the electronic charge
 and $\epsilon_{0}$ the permittivity of free space. Contrasted to the analysis of celestial circular orbits in the previous section \ref{orbit}, where we retained only quadratic powers in $|\dot{\vec{r}}|/c$, here we consider corrections till fourth power in $|\dot{\vec{r}}|/c$.
Thus the Hamiltonian of the hydrogen atom can be written from Eq.(\ref{energy}),
\begin{equation}
\label{hygrogen_ham}
\ham_{\rm rel} = \frac{p^{2}_{r}}{2 \mu \chi(r)} + \frac{L^{2}}{2 \mu r^{2}} - \frac{k}{r} +\frac{\Gamma L^2}{6 \mu r^3} + \frac{\Gamma L^4}{40 \mu c^2 r^5}  	\:	,
\end{equation} 
It is worth mentioning at this stage that the usual relativistic kinetic correction in standard texts involves expanding the relativistic kinetic energy in powers of $p/(\mu c)$ and retaining terms till the fourth power of $p$\cite{griffithExpansion}. We later argue as to why both approaches, even though retaining the same power in their respective expansions do not result in the same numerical value of the corrections. They, however match in order of magnitude, as expected. 
The $r$ dependent function $\chi (r)$ can be re-written as,
\begin{equation}
\label{chi}
\chi(r) = \left[1 + \frac{\Gamma}{r} \left(1 + \frac{L^2}{6 \mu^2 c^2 r^2} \right)  \right] = \left( 1 + \frac{\Gamma \alpha(r)}{r} \right) \: ,
\end{equation}
where we have defined $\alpha(r)$ for convenience. Readers' attention is brought to the fact that the fourth term in the relativistic Hamiltonian, $kL^2/(6 \mu^2 c^2 r^3)$ resembles the spin-orbit\cite{griffithSO} coupling term in the hydrogen atom problem. It is very interesting to note that a unified generalized potential correction to the central force Hamiltonian automatically provides a fore-runner for spin-orbit like terms, which essentially work to couple the central force with the angular momentum of the system. In particular, the spin-orbit term in the hydrogen atom involves the coupling of the electron intrinsic spin angular momentum with the orbital magnetic field of the proton seen by the electron. 
Our generalized potential approach predicts a similar form of the coupling of the central force with the total angular momentum 
of the system as a relativistic phenomenon, and may be understood as a pedagogical explanation for coupling of the spin 
and orbital angular momentum of the system with the central force. 
 Our treatment here would be semi-classical, whereby we would use the
 classical Hamiltonian derived in the earlier sections, to find
 a quantum mechanical expectation value of the first order
 correction in the ground state energy. 
 The non-relativistic, unperturbed Hamiltonian for the hydrogen atom
 $\ham _{NR}$ ($\rm NR$ stands for non-relativistic) can be easily written as,
\begin{equation}
\label{H_ham_NR}
\ham _{NR} (r, p_{r}) = \frac{p^{2} _{r}}{2 \mu} + \frac{L^{2}}{2 \mu r^{2}} - \frac{e^{2}}{4 \pi \epsilon _{0} r}  		\:	.
\end{equation}
As expected, Eq. (\ref{H_ham_NR}) corresponds to substituting
 $\Gamma = 0$ in Eq. (\ref{hygrogen_ham}). In case of the hydrogen atom,
 we see that $\Gamma \simeq 2.81 \times 10^{-15} \rm {m}$, while 
the most probable radius in the ground state is the Bohr radius
 $a_{0} (\simeq 5.29 \times 10^{-11} \rm {m})$.\cite{resnickEisbergHatom} 
 As a first approximation for the ground state, we replace $r$ in the 
 kinetic term in Eq. (\ref{hygrogen_ham}) by the Bohr radius $a_{0}$, 
 to ease the calculation. This is justified, since for the ground state 
 of the hydrogen atom, the variance in the observable corresponding to 
 $r$ is small. Since $\Gamma / a_{0} \ll 1$, we can expand
  $\chi(r)$ in the kinetic term 
 binomially and retain terms upto first order in $\Gamma/a_{0}$. The idea is to treat the 
relativistic correction to the Hamiltonian as a small perturbation
 since $\Gamma \alpha(a_{0})/a_{0} \ll 1$, and calculate the expectation value of
 the first order perturbation to the ground
 state wave energy of the hydrogen atom by treating
  $r$ and $p_{r}$ as corresponding quantum mechanical operators. The
   relativistic correction is treated as a perturbation ($\ham'$) to the non-relativistic Hamiltonian $\ham_{NR}$,
\begin{equation}
\label{Hprime}
\ham_{\rm rel} = \ham_{NR} + \ham '(r,p_{r})  	\: \: ; \: \: \ham' << \ham_{NR}	,
\end{equation}
and correspondingly the relativistic energy correction may be added to $E^{(0)}$  (the unperturbed, non-relativistic ground state energy of the hydrogen atom),
\begin{equation}
\label{energy_perty}
E = E^{(0)} + E^{(1)} _{r} \: ,
\end{equation}
where $E^{(1)} _{r}$ is the first order correction to the energy due to relativistic effects.
After replacement of $r$ by $a_{0}$ in the kinetic term and retaining upto first power in $\Gamma \alpha(a_{0})/a_{0}$, we replace the physical observable $r$ and $p_{r}$ with 
their corresponding quantum mechanical operators and write an operator
 for the perturbation in the Hamiltonian,
\begin{equation}
\label{Hprime_operator}
\hat{\ham}' (r, p_{r}) =  \left( \frac{-\Gamma \alpha(a_{0})}{a_{0}} \right) \frac{ \hat{p} ^{2} _{r}}{2 \mu} + 	\frac{\Gamma \hat{L}^2}{6 \mu\hat{ r}^3} + \frac{\Gamma \hat{L}^4}{40 \mu^3 c^2 \hat{r}^5} \:	.
\end{equation}
We use the ground state of the hydrogen atom to obtain the expectation value of the first order energy correction to the system as,
\begin{eqnarray}
\label{Er1_2}
E^{(1)} _{r} &=&  - \frac{\Gamma}{a_{0}} \left(1 + \frac{l(l+1)\hbar^2}{6 \mu^2 c^2 {a_{0}}^2 } \right) \left( E^{(0)} + \frac{e^2}{4 \pi \epsilon_{0} a_{0}}\right) \nonumber \\ &+&  \frac{\Gamma l(l+1) \hbar ^{2}}{6 \mu} \left\langle \frac{1}{\hat{r} ^{3}} \right\rangle +  \frac{\Gamma l^2(l+1)^2 \hbar ^{4}}{40 \mu^3 c^2} \left\langle \frac{1}{\hat{r} ^{5}} \right\rangle 	\:	.
\end{eqnarray}
The expectation values of powers of $(1/ \hat{r})$ can be computed from the famous Kramer's relation (or the second Pasternack relation)\cite{kramer}, \cite{kramer2}. A numerical comparison is in place here.
  The unperturbed ground state energy $E^{(0)} = -13.6 \rm {eV}$, while
   standard texts\cite{griffithsPerturbation} quote a first order 
   correction its kinetic energy solely to be $-8.99 \times 10^{-4} \rm {eV}$  by expanding the relativistic kinetic energy in powers of $p/(\mu c)$ and retaining upto fourth power in $p$. The correction due to the fine-structure which includes both kinetic and spin-orbit corrections is quoted to be $-1.81 \times 10^{-4} \rm{eV}$.
     The relativistic correction due to our Hamiltonian, using the generalized potential is found to be from Eq. (\ref{Er1_2})
      to be $-2.41 \times 10^{-4} \rm {eV}$. Thus, we see that our prediction matches more closely to the fine-structure value which includes corrections to the kinetic energy and also incorporates corrections due to angular momenta coupling to the central force. 
On utilizing the Schr\"{o}dinger's Equation to find out the expectation value of $p^4$ in the standard approach\footnote{In the standard approach, the first-order energy correction is proportional to $<p^4>$ and is given by, 
\begin{eqnarray}
E^{(1)} _{r} &=& -\frac{1}{2 m c^2} \left\langle (E-V)^2 \right\rangle \:  \nonumber \\ &=& \: -\frac{1}{2 m c^2} \left[ E^{2} _{n} + 2E_{n} \frac{e^2}{4 \pi \epsilon_{0}} \left\langle \frac{1}{\hat{r}} \right\rangle + \left(\frac{e^2}{4 \pi \epsilon_{0}}\right)^{2} \left\langle \frac{1}{\hat{r} ^2} \right\rangle \right]
\end{eqnarray}
}, we notice that it splits into terms containing $1/r$ and $1/r^2$. This can also be seen in the kinetic correction term (first term) in our Hamiltonian of Eq. (\ref{hygrogen_ham}), which when expanded in powers of $\Gamma/r$ yields similar expressions as in the standard textbook approach. Thus, we stress on the similarity of structure of the kinetic energy corrections in both the approaches. Our methodology of building a generalized potential energy is able to give both a kinetic correction and in addition generates terms which correct the potential energy of the system as well. Corrections to the potential energy of a system due to relativistic effects are not very surprising. A famous motivation for their existence is the Darwin term\cite{darwin} which occurs naturally in Dirac equation and involves corrections to the effective potential at the nucleus in the hydrogen atom. Our Hamiltonian splits into terms independent of $\dot{r}$ and also terms depending on various powers of $\dot{r}$. It allows for coupling of the central force with the angular momentum of the system (spin and orbital) and produces corresponding corrections to the system, which match closely to the predicted value of the fine-structure corrections. Our unified methodology which aims to reproduce the force equation using a suitable generalized potential energy and consequently study the corrections it produces in the Hamiltonian. Thus, our approach is different from the regular kinetic correction in texts and leads to a different Hamiltonian which adds additional correction terms (to the potential energy) due to coupling of the angular momentum of the system with the central force and hence the two numerical values are not expected to match exactly. 
Hence, using a generalized potential and a semi-classical approach, we are able to predict the correct order of the correction in the energy due to a relativistic correction, which can be seen as a model application for the relativistically corrected Hamiltonian.
\section{Conclusion} \label{conclude}
In our novel approach to introduce relativistic corrections
 to the central force problem, we have derived a generalized
 potential energy function which depends both on the position and velocity
 of the two-body system. The Lagrangian constructed has a familiar
 classical form, only with the replacement of the ordinary potential
 with the generalized counterpart derived in the paper. Our methodology employs a power series expansion in all powers in $|\vec{v}|/c$ for the generalized potential energy. The new Lagrangian
 is able to reproduce the relativistic force equation for the particle under the influence of the central force. Thus, we did not have to indulge in a covariant
 formulation (Lorentz invariance), while we used
 a simplistic Lagrangian methodology to extract the basic physics of the
 problem at hand. The Hamiltonian of the system can easily be constructed and used to calculate relativistic corrections to the total energy of the system. 
We employ our model to derive 
and understand the relativistic corrections in a central force, {\em namely} 
(i) to the circular orbits in celestial mechanics and (ii) to the ground state 
Bohr energy of the hydrogen atom and found a reasonable match with known 
results.
\section*{Acknowledgements}
The authors would like to thank the anonymous reviewers and the Associate Editor of the journal for their constructive comments to help improve the manuscript.

\end{document}